\documentclass[prl,twocolumn,aps,showpacs]{revtex4}
\pdfoutput=1

\usepackage{graphicx}

\begin{document}

\title{Topological Crystalline Kondo Insulator in Mixed Valence
  Ytterbium Borides}

\author{Hongming Weng, Jianzhou Zhao, Zhijun Wang, Zhong Fang and Xi
  Dai}

\affiliation{ Beijing National Laboratory for Condensed Matter Physics,
  and Institute of Physics, Chinese Academy of Sciences,
   Beijing 100190, China }

\date{\today}

\begin{abstract}
  The electronic structures of two mixed valence insulators YbB$_6$
  and YbB$_{12}$ are studied by using the local density approximation
  (LDA) supplemented with the Gutzwiller  method and dynamic mean field
  theory (DMFT). YbB$_6$ is found to be a moderately correlated
  $Z_2$ topological insulator, similar to SmB$_6$ but having much
  larger bulk band gap. Notably, YbB$_{12}$ is revealed to be in a new novel 
  quantum state, strongly correlated topological crystalline Kondo insulator, 
  which is characterized by its non-zero mirror Chern number. 
  The surface calculations find odd (three) and even
  (four) number of Dirac cones for YbB$_6$ and YbB$_{12}$,
  respectively.

\end{abstract}

\maketitle

Topological Insulators (TI)~\cite{Hasan:2010vc,Qi:2011wt} have been
extensively studied, but mostly on the $s$ and $p$ orbital systems,
such as HgTe~\cite{Bernevig:2006ij,HgTe:exp,Dai:2008}, and
Bi$_2$Se$_3$ family
compounds~\cite{Zhang:2009ks,Xia:2009km,ChenYL:Bi2Te3}, which are free
of strong correlation effects. In the presence of strong electron
interactions, much fruitful topological phases might be expected, such
as the topological Mott~\cite{Pesin:2010} or
Kondo~\cite{Dzero:2010dj,Dzero:2012kx,Lu:2013hi} insulators,
topological superconductors~\cite{Qi:2011wt}, and fractional
TI~\cite{Sheng:2011iv,Regnault:2011}. To pursue those exotic phases,
however, an important and necessary step is to find suitable
compounds, which are strongly correlated (presumably in $d$ and $f$
orbital systems) and topologically non-trivial. Studies on such
systems are challenging both theoretically and experimentally.
Nevertheless, the mixed valence phenomena provides an important way
towards this direction~\cite{Martin:1979tf,Coleman:2007,Lu:2013hi}. For instance, in rare-earth
mixed valence compounds, the band inversion naturally happens between
the correlated 4$f$ and 5$d$ states, which may lead to correlated
topological phases. SmB$_6$, a typical mixed valence compound, has
been proposed theoretically as ``topological Kondo
insulator''~\cite{Dzero:2010dj,Dzero:2012kx,Lu:2013hi}, and recently
been supported by
transport~\cite{Wolgast:2012ww,Kim:2012dj,Thoma:2013st,Li:2013wl} ,
photo emission~\cite{Neupane:2013wt,Jiang:2013tl,Xu:2013th} and
STM~\cite{Yee:2013ve} experiments. Here the dispersive 5$d$ conduction
band intersects with the $4f$ energy levels, leading to electron
transfer and strong quantum fluctuation among 4$f$ atomic
configurations~\cite{Coleman:2007}. At sufficiently low temperature,
the coherent motion of 4$f$ states are established, resulting in the
formation of ``heavy fermion'' bands, whose non-trivial $Z_2$
topological index~\cite{Hasan:2010vc,Qi:2011wt} can be determined by
the single particle Green's function at zero
frequency~\cite{Wang:2010tg,Wang:2012fj}.

In the present paper, we will focus on another family of binary mixed
valence compounds, Ytterbium Borides, and proposed that various
correlated toplogical phases, in particular a new topological
crystalline Kondo insulator~\cite{Teo:2008bk,
  Hsieh:2012cn,Sun:2013tckl}, can be realized. Among the four typical
compounds, YbB$_4$, YbB$_6$, YbB$_{12}$ and YbB$_{66}$, the Yb ions in
YbB$_4$ and YbB$_{66}$ are in $2+$ or $3+$ respectively, while the XPS
and XAS data suggest that the valence of Yb in YbB$_6$ and YbB$_{12}$
is around $2.2$~\cite{Nanba1993557} and
$2.8$~\cite{Kasaya1983437,Kasaya1985429,Susaki:1996vb,Takeda:2004iq}
respectively, indicating the mixed valence nature. As we have proposed
in reference~\cite{Lu:2013hi}, the local density approxiamtion (LDA)
combined with the Gutzwiller density functional
theory~\cite{Deng:2009p6949} is a powerful tool to compute the ground
state and the quasi-particle spectrum of such correlated
systems. Using this method, we find that (1) YbB$_6$ is a correlated
$Z_2$ topological insulator similar to SmB$_6$ but with much larger
band gap (31 vs 10 meV); (2) YbB$_{12}$ is a new topological
crystalline Kondo insulator, which can be characterized by the
non-trivial mirror Chern number~\cite{Teo:2008bk, Hsieh:2012cn}, and
shows even number of Dirac cones on its surface.

 \begin{figure}
   \includegraphics[width=0.45\textwidth]{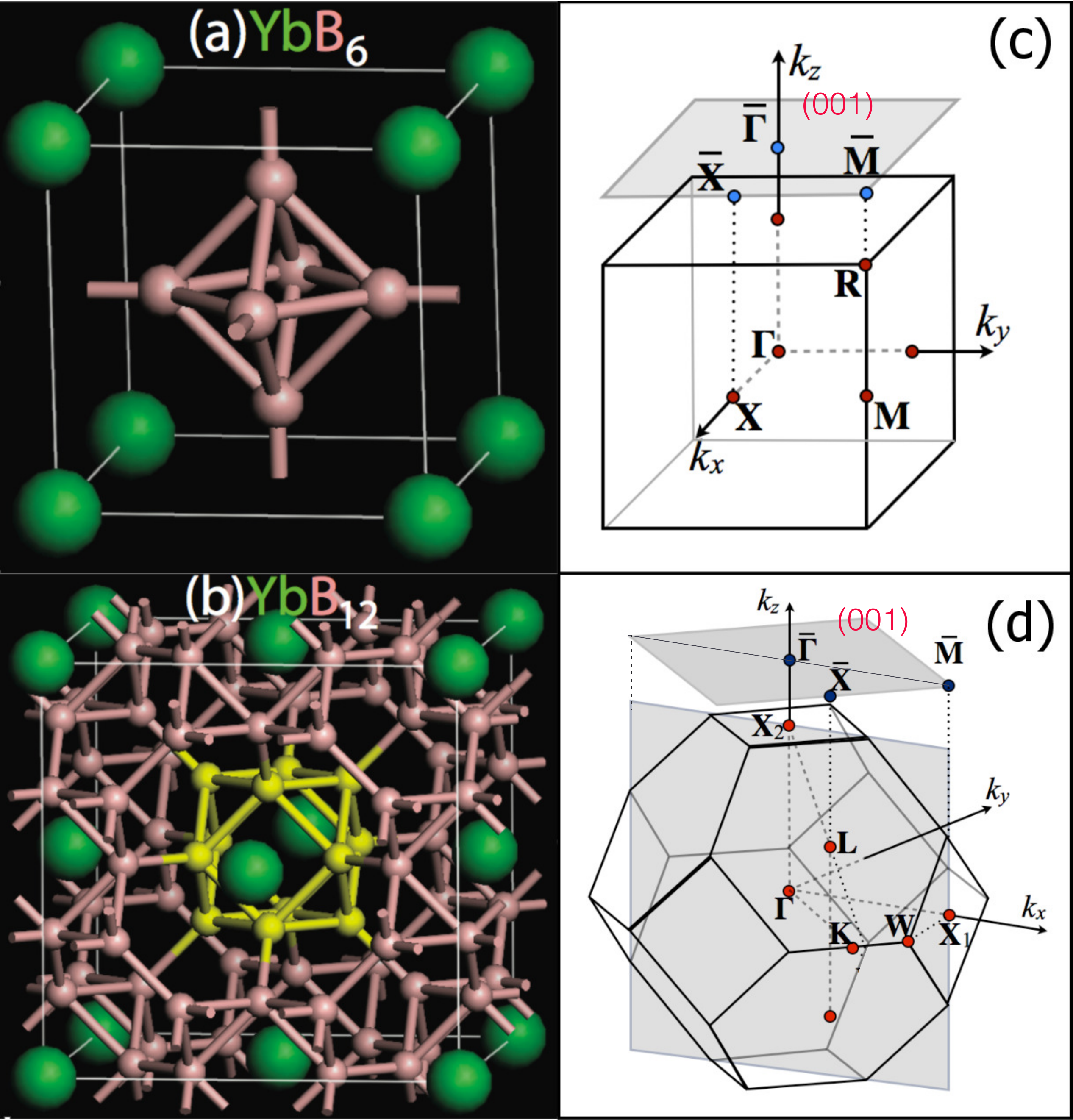}
   \caption{(a) The CsCl-type structure of YbB$_6$ with $Pm\bar{3}m$
     space group, and (b) the NaCl-type structure of YbB$_{12}$ with
     $Fm\bar{3}m$ space group (B$_{12}$ cubo-octahedral cluster is
     highlighted). (c) and (d) are the corresponding bulk and surface Brillouin
     Zones.}\label{crystal} \end{figure}

 As shown in Fig.\ref{crystal}, YbB$_6$ has the CsCl-type structure,
 the same as SmB$_6$, with Yb and B$_6$ octahedral cluster occupying
 Cs and Cl site, respectively; while YbB$_{12}$ takes the NaCl-type
 structure with Yb and B$_{12}$ cubo-octahedral cluster replacing the
 Na and Cl ions, respectively. The LDA part of the calculations have
 been done by full potential linearized augmented plane wave method
 implemented in the WIEN2k package~\cite{Blaha:2001vw}. A regular mesh
 of 12$\times$12$\times$12 $k$ points is used, and the muffin-tin
 radii ($R_{MT}$) of Yb and B atoms are taken as 2.50 and 1.57
 bohr. The plane-wave cutoff $K_{max}$ is given by $R_{MT}K_{max}$=$
 7.0$. The spin-orbit coupling (SOC) is included self-consistently in
 all calculations.

\begin{figure}
\includegraphics[width=0.45\textwidth]{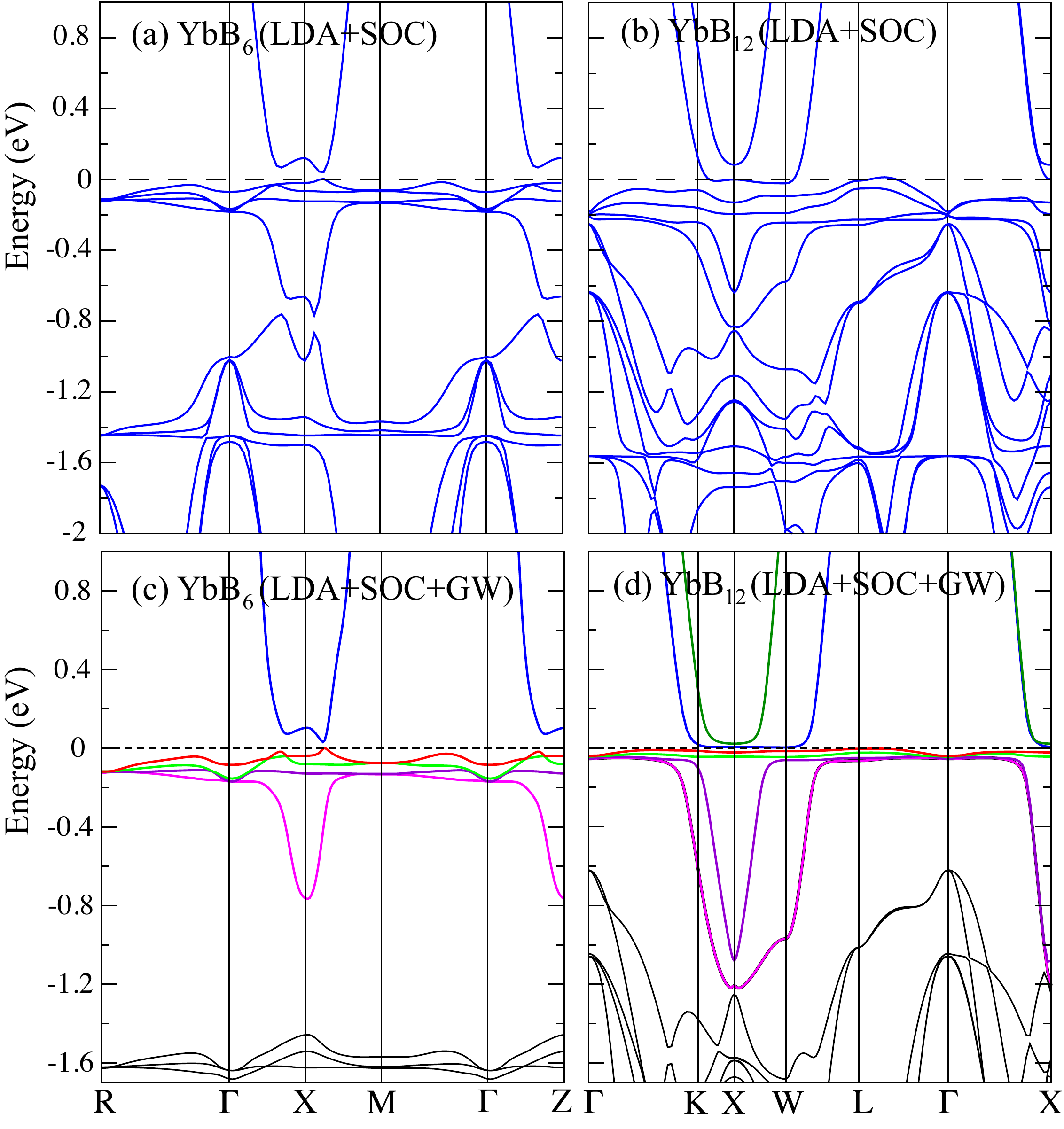}
\caption{ The band structure of (a) YbB$_6$ and (b) YbB$_{12}$ obtained from LDA+SOC
  calculations. (c) and (d) are their quasi-particle band structures
  calculated from LDA+SOC+Gutzwiller with $U$=6.0 eV. } \label{band}
\end{figure}

The LDA band structures, shown in Fig.\ref{band} (a) and (b), suggest
that the major features are very similar to SmB$_6$. Firstly, the
Yb-$4f$ orbitals, which split into the $j$=5/2 and $j$=7/2 manifolds
due to the SOC, form two sets of narrow bands with the former fully
occupied and the later near the Fermi level (in SmB$_6$, the $j$=7/2
manifolds are fully empty, and $j$=5/2 states are close to the Fermi
level). Secondly, the low energy band structure is semiconducting with
a minimum gap of about 29 meV along the $X$-$M$ path in YbB$_6$, and a
nearly zero indirect gap for YbB$_{12}$. Thirdly, there are clear band
inversion features around the $X$  point in both systems. In YbB$_6$,
one $5d$ band goes below the $j$=7/2 bands (by about 1.0eV), which
reduces the occupation number $n_f$ of the 4$f$-states to be around
13.58 (resulting in the Yb valence of +2.42). What is qualitatively
different in YbB$_{12}$ is that two 5$d$ bands (which strongly
hybridize with B-2$s$ and -2$p$ states) sink down below the $j$=7/2
states (by about 0.8eV), and $n_f$ is further reduced to be 13.31,
leading to the Yb valence of +2.69. We noticed that the shortest Yb-B
bond length in YbB$_{12}$ (2.277 \AA) is much shorter than that in
YbB$_6$ (by about 0.772 \AA), the enhanced 5$d$-2$p$ hybridization in
YB$_{12}$ therefore push one more 5$d$ state down to be lower than the
4$f$ states at X point.  As has been discussed in
SmB$_6$~\cite{Lu:2013hi}, the hybridization between the 5$d$ and 4$f$
states will open up a gap, and generate the semiconducting behavior.
Since the 5$d$ and the 4$f$ states have opposite parity at the $X$
point, and further more there are three $X$ points in the whole BZ,
the band inversion in YbB$_6$ happens three (odd) times, which leads
to a non-trivial TI with the $Z_2$ indices given as
$(1;111)$~\cite{Fu:2007io,Fu:2007ei}.  While for YbB$_{12}$, the two
times of band inversion at each $X$ point generates totally six (even)
times band inversion in the whole BZ, which gives a trivial insulator
in the sense of $Z_2$=0.

\begin{table}
\caption{The products of parity eigenvalues of the occupied 
states for TRIM points, $\Gamma$, $X$, $R$, and $M$ for YbB$_6$
and $\Gamma$, $X$, $L$, and $L$ for YbB$_{12}$ in the BZ. 
$n_f$ is the occupation number of $4f$ orbitals by LDA+Gutzwiller, 
compared with LDA results in the brackets and the experimental one $n^{exp}_{f}$. And $z$ is the quasi-particle 
weight obtained by LDA+Gutzwiller. }\label{parity}
\begin{tabular*}{0.50\textwidth}{@{\extracolsep{\fill}}cccccccc}
  	\hline\hline
      & $\Gamma$ & 3$X$  & 3$M$($L$) & $R$($L$) & $n_f$ & $n^{exp}_{f}$ & $z$  \\
  	\hline
  	YbB$_6$  &  + &  - & + & +  & 13.80(13.58) & 13.8\footnote{Ref. \cite{Nanba1993557}} & 0.87 \\
  	YbB$_{12}$ &  + & + & + & + & 13.11(13.31) & 13.14\footnote{Ref. \cite{Susaki:1996vb}}, 13.12\footnote{Ref. \cite{Takeda:2004iq}} & 0.28 \\
  	\hline\hline
\end{tabular*}
\end{table}

Due to the partially filled $4f$ states near the Fermi level, the
on-site interactions among the $f$-electrons are expected to play
important roles, which can be captured by the LDA+Gutzwiller
method~\cite{Deng:2009p6949,Lu:2013hi}.  For both systems, we take the
Hubbard interaction $U$ of 6.0 eV and neglect the Hund's
coupling $J$ .
 From the calculated results (shown in Fig.2(c) and (d)), we find
three major modifications coming from the correlation effects.
Firstly, the 4$f$ occupation number is further pushed towards its
integer limit, namely towards $n_f\sim 13.0$ for YbB$_{12}$ and
$n_f\sim 14.0$ for YbB$_6$ respectively, being in better agreement
with the experimental data (see Table I). This is simply due to the fact that the
strong Coulomb interaction tend to suppress the charge fluctuation
among different atomic configurations. 
 Secondly, we find renormalization of the 4$f$
quasi-particle bands, and the behaviors of YbB$_6$ and YbB$_{12}$ are
quite different. The quasi-particle weight $z$ is 0.87 for YbB$_6$ but
reaches very low 0.28 for YbB$_{12}$, indicating that the former is an
intermediately correlated insulator while the latter is very close to
the strong coupling description, the Kondo insulator. Since the same
interaction parameters are used for both materials, the big difference
in $z$ is due to the different 4$f$ occupations $n_f$.  Thirdly, we
find that the hybridization gap between 4$f$ and itinerant 5$d$ bands
is slightly enlarged to be 31 meV in YbB$_6$ and 6 meV in YbB$_{12}$,
being much closer to the experimental
values\cite{Werheit:2000er,Takeda:2006bb,Susaki:1999vg}.  The parity
analysis is still applicable for the quasi-particle bands obtained by
LDA+Gutzwiller~\cite{Lu:2013hi,Wang:2010tg,Wang:2012fj}, and the
results listed in Table I conclude that $Z_2$ indices keep unchanged
after including the correlation effects for both materials.


\begin{figure}
\includegraphics[width=0.45\textwidth]{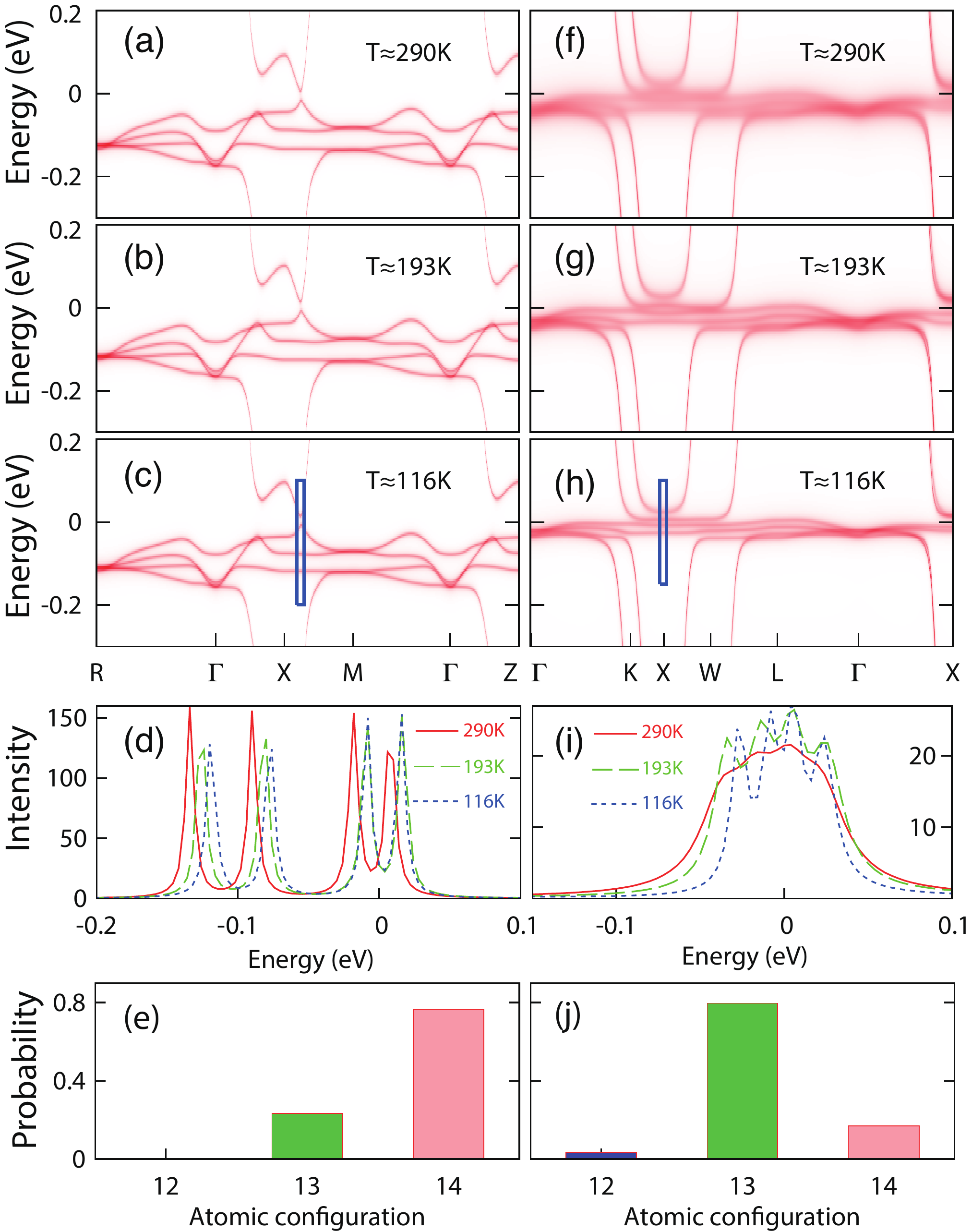}
\caption{The momentum-resolved spectral function $A_k(\omega)$ of
  YbB$_6$ (a - c) and YbB$_{12}$ (e - g) at different temperature
  ($T\approx 290$K, $193$K, $116$K from top to bottom). (d) YbB$_6$
  spectral function $A(\omega)$ at the k 
  point with the minimum gap as indicate in (c) with $T\approx$ 290K(red),
  193K(green), 116K(blue). (i) YbB$_{12}$ spectral function
  $A(\omega)$ at the $X$ point, as indicate in
  (h), with $T\approx$ 290K(red), 193K(green), 116K(blue). (e) and (j)
  are the probability of atomic eigenstates with occupation number
  $N_f = 12, 13, 14$ obtained by LDA+DMFT for YbB$_6$ and YbB$_{12}$
  at $T \approx 116$K, respectively. }\label{finite_temp}
\end{figure}

One of the major differences between Kondo insulator and band
insulator is the temperature dependence of electronic structures. For
a typical band insulator, the band picture is applicable for almost
all temperature range and the rigid band approximation is usually
adopted; while for a Kondo insulator, the coherent hybridization
between the localized $f$ orbitals and the conduction bands (leading
to insulating behavior) only occurs below the Kondo temperature, which
has been found to be around 220K for YbB$_{12}$
\cite{Susaki:1996vb}. In order to calculate the electronic structure
at finite temperature, we further apply the LDA+DMFT (dynamical mean
field theory) method \cite{Georges:1996hv,Kotliar:2006p6774} to both
materials. We use the continuous time quantum Monte Carlo method based
on the hybridization expansion \cite{Gull:2011jd} for the impurity
solver of DMFT, and take the same interaction parameters. The
electronic spectral functions (shown in Fig.\ref{finite_temp})
obtained by the maximum entropy method \cite{Haule:2010p6732} suggest
that two materials behave quite differently. At low temperature
($T$=116K), the spectral functions for both materials are in good
agreement with the LDA+Gutzwiller results (plotted in Fig.\ref{band}).
At 290K, however, the spectral function of YbB$_{12}$ is significantly
smeared out, while that of YbB${_6}$ still keeps unchanged. This
indicates that the rigid band picture is applicable to YbB$_{6}$ but
broken down for YbB$_{12}$, which can be viewed as a Kondo Insulator
with Kondo temperature around $200$K.


\begin{figure}
\includegraphics[width=0.45\textwidth]{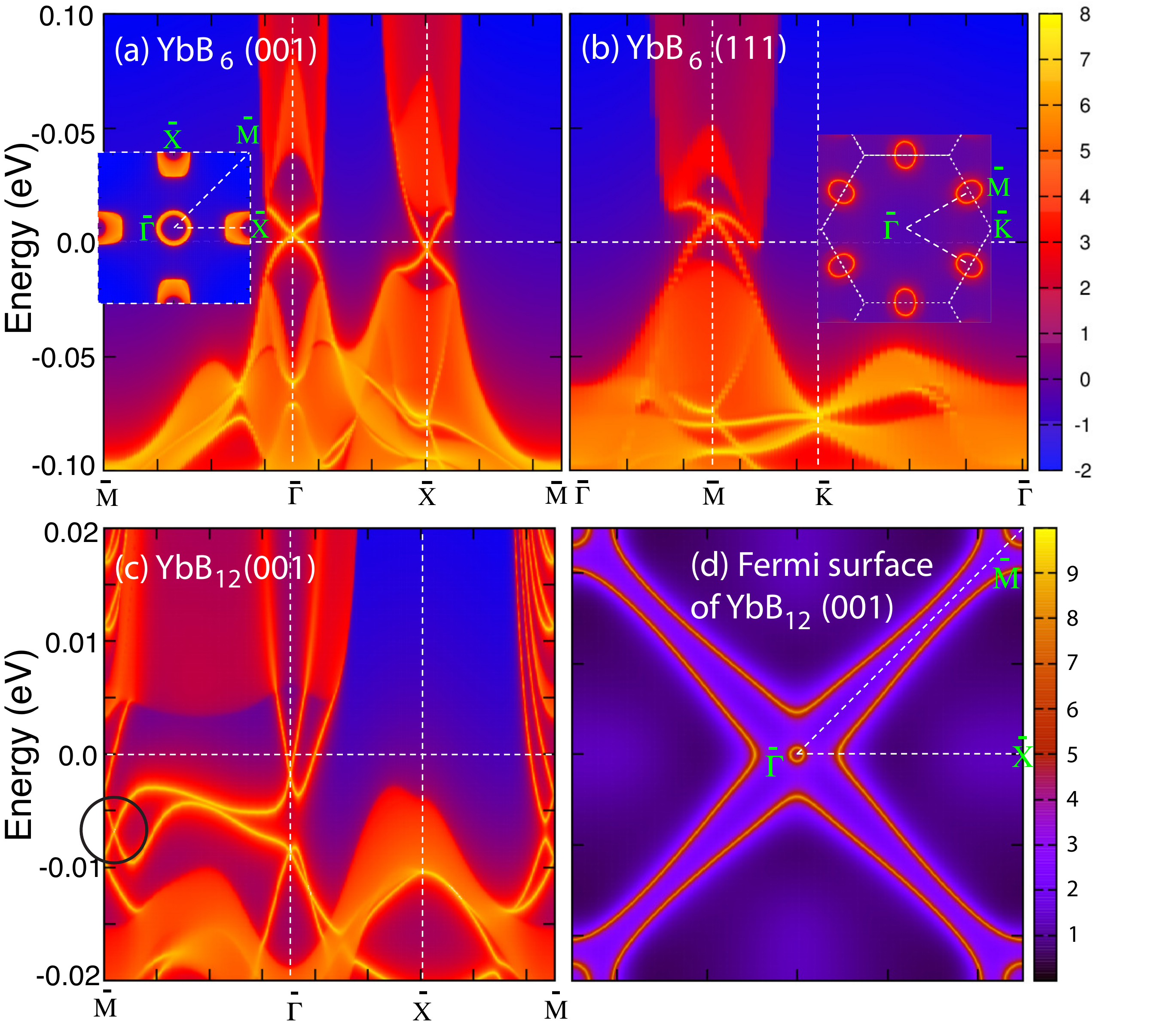}
\caption{The surface states (SS) of  YbB$_6$ for its (a) (001) and (b)
  (111) surface from LDA+Gutzwiller calculation. Insets are the Fermi
  surfaces with chemical potential 5 meV above and below the Dirac
  point at $\bar{\Gamma}$ drawn for (001) and (111) SS,
  respectively. (c) SS of YbB$_{12}$ (001) surface from LDA+Gutzwiller
  calculation and (d) its Fermi surface at Fermi level. The Dirac cone 
 due to nonzero MCN is indicated by a circle in (c).}
\label{surface}
\end{figure}


The surface states (SS) for YbB$_6$ and YbB$_{12}$ (shown in
Fig.~\ref{surface}) are obtained by using the Green's function method
based on the tight-binding model constructed from the Maximally
Localized Wannier Functions. The correlation corrections from
Gutzwiller approximation are included.  For YbB$_6$, the SS on (001)
surface is very similar to that of SmB$_6$~\cite{Lu:2013hi}, which
contains three surface Dirac cones located at $\bar{\Gamma}$ and two
$\bar{X}$ points; for its (111) surface we also find three Dirac cones
located at $\bar M$, which are symmetric due to the three fold
rotation along [111] axis.  Although YbB$_{12}$ is topologically
trivial in the sense of $Z_2$=0, the band inversion feature around the
$X$ points generates surfaces states as well, which are shown in
Fig.\ref{surface}(c) and (d). Unlike YbB$_6$, whose SS has odd number
of Dirac points, the SS of YbB$_{12}$ contains four Dirac points on
the (001) surfaces (near the $\bar M$ point along the $\bar M$ to
$\bar \Gamma$ direction), indicating that it is a topological
crystalline insulator similar to
SnTe~\cite{Teo:2008bk,Hsieh:2012cn,Ando:SnTe,Dziawa:2012ph}.  The even
number of Dirac points are protected by the reflection symmetry
respect to the (100) or (010) planes (i.e., the $\Gamma X_1 X_2$ plane
in Fig. 1(d)), and is the consequence of non-zero ``Mirror
Chern number'' (MCN) within such planes, which can be defined as the
Chern number of half of the occupied states (distinguished by the
different eigenvalues of mirror
symmetry)~\cite{Teo:2008bk,Hsieh:2012cn}.  We apply the Wilson loop
method introduced in reference~\cite{Yu:2011dx} to calculate the MCN
of YbB$_{12}$, and get MCN $=2$ for the $\Gamma X_1 X_2$ plane, which
is consistent with the SS behavior observed on $(001)$ surface. In
fact the nonzero MCN obtained for the $(100)$ mirror plane implies the
appearance of SS on any surface with index $(0nm)$.  Unlike the
situation in SnTe, however, here the possible topological SS can
appear only below the Kondo temperature, when the local 4$f$ moments
are effectively screened by the conduction bands and the heavy
quasi-particles form.  Therefore the nonzero MCN in YbB$_{12}$
indicates that the ground state of YbB$_{12}$ is a new topological
crystalline Kondo insulator~\cite{Sun:2013tckl}.

In summary, we have applied the LDA+Gutzwiller and LDA+DMFT methods to
study the possible correlated topological phases in two mixed valence
Yb compounds YbB$_6$ and YbB$_{12}$. Our results verify that YbB$_6$
is a moderately correlated $Z_2$ topological insulator, while
YbB$_{12}$ is a strongly correlated topological crystalline Kondo
insulator with MCN=2.  This work was supported by the NSF of China and
by the 973 program of China (No. 2011CBA00108 and 2013CBP21700). We
acknowledge the helpful discussions with professor P. Coleman
and Yulin Chen.


\bibliography{YbB6YbB12_v3}


\end{document}